\documentclass[traditabstract]{aa}
\usepackage{graphicx,natbib,txfonts}
\bibpunct{(}{)}{;}{a}{}{,}

\begin{document}

\title{The short GRB\,070707 afterglow and its very faint host galaxy\thanks{Based on observations made with  ESO Telescopes
   at the La Silla Paranal Observatory under program ID 079.D-0909.}}

\author{S. Piranomonte\inst{1} \and
P. D'Avanzo\inst{2} \and
S. Covino\inst{2} \and
L. A. Antonelli\inst{1} \and
A. P. Beardmore\inst{3} \and
S. Campana\inst{2} \and
G. Chincarini\inst{2,4} \and
V. D'Elia\inst{1} \and
M. Della Valle\inst{5,6,7} \and
F. Fiore\inst{1} \and
D. Fugazza\inst{2} \and
D. Guetta\inst{1} \and
C. Guidorzi\inst{2} \and
G. L. Israel\inst{1} \and
D. Lazzati\inst{8,9} \and
D. Malesani\inst{10} \and
A. M. Parsons\inst{11} \and
R. Perna\inst{8} \and
L. Stella\inst{1} \and
G. Tagliaferri\inst{2} \and
S. D. Vergani\inst{12,13}
}

\offprints{S. Piranomonte, \email{piranomonte@oa-roma.inaf.it}}

\institute{INAF-Osservatorio Astronomico di Roma, via Frascati 33, Monteporzio-Catone (RM), 00040 Italy. \and
INAF-Osservatorio Astronomico di Brera, via E. Bianchi 46, 23807 Merate (LC), Italy. \and
Department of Physics and Astronomy, University of Leicester, Leicester, UK. \and
Universit\`a degli Studi di Milano-Bicocca, Dipartimento di Fisica, piazza delle Scienze 3, 20126 Milano, Italy. \and
INAF-Osservatorio Astronomico di Capodimonte, salita Moiariello 16, 80131 Napoli, Italy. \and
European Southern Observatory, Karl-Schwarzschild-Strasse 2, D-85748 Garching bei M\"unchen, Germany \and
International Center for Relativistic Astrophysics Network, P.le della Repubblica, 10, I-65122, Pescara, Italy \and
JILA, University of Colorado, Boulder, CO 80309-0440, USA. \and
Department of Physics, North Carolina State University, Box 8202,Raleigh, NC 27695 \and
Dark Cosmology Centre, Niels Bohr Instute, University of Copenhagen, Juliane Maries vej 30, 2100 K\o{}benavn \O, Denmark. \and
NASA/Goddard Space Flight Center, Greenbelt, Maryland 20771, USA. \and
Dunsink Observatory, DIAS, Dunsink lane, Dublin 15, Ireland. \and
School of Physical Sciences and NCPST, Dublin City University, Dublin 9, Ireland.
}

\date{Received <date> / Accepted <date>}

\abstract{
We present the results from an ESO/VLT campaign aimed at studying the afterglow
properties of the short/hard gamma ray burst GRB\,070707. Observations were
carried out at ten different epochs from $\sim 0.5$ to $\sim 80$ days after
the event. The optical flux decayed steeply with a power-law decay index
greater than 3, later levelling off at $R\sim 27.3$~mag; this is likely the
emission level of the host galaxy, the faintest yet detected for a short GRB.
Spectroscopic observations did not reveal any line features/edges that could
unambiguously pinpoint the GRB redshift, but set a limit $z < 3.6$.
In the range of allowed redshifts, the host has a low luminosity, comparable
to that of long-duration GRBs. The existence of such faint host galaxies
suggests caution when associating short GRBs with bright, offset galaxies,
where the true host might just be too dim for detection. The steepness of the
decay of the optical afterglow of GRB\,070707 challenges external shock models
for the optical afterglow of short/hard GRBs. We argue that this behaviour
might results from prolonged activity of the central engine or require alternative scenarios.}

\keywords{gamma rays: bursts }


\authorrunning {Piranomonte et al.}  
\titlerunning{The afterglow of \object{GRB\,070707}}

\maketitle


\section{Introduction}

Gamma-ray bursts (GRBs) are among the most powerful explosions in the universe.
They are revealed in the hard X-ray/soft gamma-ray band and are followed in
many cases by a fading afterglow observable from radio to X-ray wavelengths.
GRBs are empirically classified in two groups \citep{Maz81, Nor84, Kou93,Tav96}: short GRBs
last less than 2\,s and have a hard spectrum; long GRBs have longer durations
(typically tens to hundreds seconds) and somewhat softer spectra.  The
emergence of a typical supernova (SN) spectrum superposed on the rapidly
decaying non-thermal afterglow weeks after the events and the association with
blue, highly star-forming galaxies provided strong evidence that a significant
fraction of long GRBs originates in the gravitational collapse of massive
stars (but see \citealt{MDV06, Fyn06, Gal06a}).

Short GRBs are revealed less frequently than long GRBs \citep[they comprise 
about 1/4 and 1/10 of the  BATSE and \textit{Swift} samples,
respectively;][]{Kou93,Ber07}; moreover their afterglows are weaker and, thus,
more difficult to detect and follow up. These are among the reasons why the 
origin of short GRBs is still under debate, despite the important progresses
made in the \textit{Swift} era. The tight upper limits on any associated SN
\citep{Hjo05,Cov06,Kann08} as well as the association with a broad variety of 
Hubble types hosts, from elliptical \citep{Berger05} to moderate star forming galaxies \citep[e.g.][]{Cov06, Fox06, Berger08},
rules out the core-collapse mechanism as the main channel for short-GRBs
production and strongly suggest that the explosion mechanism and/or progenitors
of short GRBs are different from those of long GRBs \citep[for a recent review,
see][]{LeRa07}.

The leading model for short GRBs involves the merging of a system composed of
two collapsed objects, a double neutron star (DNS) or a black hole/neutron star
binary. In those systems that evolve out of  massive stars that were born in a
binary system (we term these ``primordial binaries''), the delay between
formation and merging is dominated by the gravitational wave inspiral time,
ranging from tens of Myr to a few Gyr \citep{PeBe02}, strongly dependent on the
initial system separation. 
Short GRBs that result from them are expected to: (a) have a redshift
distribution which broadly follows that of star formation 
and (b) drift away in some cases from the star-forming
regions in which they were born, and merge outside, or in the outskirts, of
galaxies \citep{Bel02}. 

A fraction of two collapsed object binaries may also form dynamically through
binary exchange interactions in the core of globular clusters \citep{Gri06}.
For such a formation mechanism, the delay between star-formation and merging is
driven by the cluster core collapse time, which is comparable to the
Hubble time \citep{Hop06}. Therefore, short GRBs originating from dynamically
formed double collapsed object binaries should go off at lower redshifts than
short GRBs from primordial binaries \citep{GuPi05,GuPi06,Gal06,Hop06,Salva08}.
In another scenario, a fraction of the short GRBs is due to hyperflares from
soft gamma-ray repeaters in the local  universe (distances up to $\sim
100$\,Mpc; \citealt{Hur05,Tan05,Fre07,Maz08}).

The above summary emphasises that redshift determination, GRB position relative
to the host galaxy and properties of the host galaxy are all crucial pieces of
information for understanding short GRBs and discriminating among different
models \citep{Bel06}. One key issue in the study of short GRBs  is the secure
identification of the host galaxy. Indeed, several short GRBs afterglows have
been detected only in X-rays and thus localized with a precision of a few
arcseconds (whereas sub-arcesecond localizations are required to unambiguously
reveal their host). The possibility that  a sizeable fraction of short GRBs lie
outside the light of their hosts, as predicted for both primordial and
dynamically formed NS-NS/BH binaries, further complicates the identification
process. Indeed some proposed short GRB associations with bright, nearby
galaxies, based on some angular separation,   might result from by-chance
alignment.


So far, a dozen short GRBs were localized with sub-arcsecond precision and host
galaxies were firmly detected with small offset. Only in a few cases (e.g.
\object{GRB\,061201}: \citealt{Str07}; \object{GRB\,080503}:
\citealt{Perley08}) no host galaxy was found down to $R \geq 26$--28 after the
optical afterglow had faded.
For other short GRBs at unknown redshift, a putative host galaxy was proposed
with magnitude $R \sim 23$--26. Spectroscopy of the brightest four of these
galaxies indicates that they lie at $0.4 \leq z \leq 1.1$. A comparison with
field galaxy magnitudes suggests that the rest of the sample lies at $z \geq 1$
\citep{Ber07b}. The unambiguously localized hosts are both early- and late-type
galaxies, with very different star formation rates and masses \citep{Nak07}.

The association with early-type galaxies has provided clear evidence that at
least a fraction of the short GRBs have progenitors related to an older stellar
population than that of long GRBs, as expected from the compact binary system
models. The nature of the progenitors of short GRBs that go off in star forming
galaxies is still under debate. The possibility that short GRBs comprise
different subclasses cannot be confirmed yet but neither excluded. 

In this paper we present the results from an extensive campaign aimed at
studying the optical afterglow of \object{GRB\,070707}. This campaign monitored
the decay of the optical afterglow evolution from about 0.45\,days to more than
one month after the burst.
This is one of the best optical light curves for a short/hard GRB so far
obtained. In Sect.\,\ref{sec:grb} we report on previous results on 
\object{GRB\,070707}. In Sect.\,\ref{sec:obs} our observations and data
analyses are discussed and in Sect.\,\ref{sec:res} our results are presented.
In Sect.\,\ref{sec:disc} we discuss our findings. 
 
\begin{table*}
\caption{VLT observation log for GRB\,070707. Magnitudes are not corrected for
Galactic absorption. Upper limits are given at $3\sigma$ confidence level.}
\centering          
\begin{tabular}{llrcccc}
\hline
Mean time          &  Exposure time             & $t-t_0$    &  Seeing     & Instrument &  Magnitude   & Filter / Grism \\
(UT)               &  (s)                       &  (days)    & ($\arcsec$) &            &              &                \\ 
\hline
2007 Jul  08.12988 &  $10 \times 120$           & 0.45742    & 0.9  & VLT/FORS1 & $23.05 \pm 0.02$   & $R$       \\
2007 Jul  09.07984 &  $10 \times 120$           & 1.40738    & 1.0  & VLT/FORS1 & $23.86 \pm 0.05$   & $R$       \\
2007 Jul  09.23005 &  $ 1 \times 180$           & 1.55759    & 1.0  & VLT/FORS1 & $24.07 \pm 0.12$   & $R$       \\
2007 Jul  10.14491 &  $10 \times 120$           & 2.47245    & 0.7  & VLT/FORS1 & $25.33 \pm 0.08$   & $R$       \\
2007 Jul  10.21523 &  $20 \times 3 \times 30$   & 2.54277    & 0.7  & VLT/ISAAC & $> 23.6        $   & $J$       \\
2007 Jul  11.13697 &  $20  \times 180$          & 3.46721    & 0.5  & VLT/FORS1 & $26.62 \pm 0.18$   & $R$       \\
2007 Jul  12.17370 &  $17 \times 300$           & 4.50124    & 0.8  & VLT/FORS1 & $26.81 \pm 0.22$   & $R$       \\
2007 Jul  15.21785 &  $18 \times 300$           & 7.54539    & 0.6  & VLT/FORS1 & $27.39 \pm 0.22$   & $R$       \\
2007 Jul  19.12752 &  $20 \times 300$           & 11.45506   & 0.6  & VLT/FORS1 & $27.21 \pm 0.20$   & $R$       \\
2007 Aug  15.07151 &  $20 \times 300$           & 38.38355   & 0.6  & VLT/FORS1 & $27.15 \pm 0.29$   & $R$       \\
2007 Sep  28.03125 &  $20 \times 3 \times 60$   & 81.95974   & 0.5  & VLT/NACO  & $> 22.7$           & $K$       \\
\hline
2007 Jul  09.32508 &  $5 \times 2400$           & 1.65262    & 1.0  & VLT/FORS1 & ---       & $300V+$GG375   \\
2007 Jul  13.15960 &  $2 \times 900$            & 5.48714    & 0.8  & VLT/FORS1 & ---       & $300V+$GG375   \\
\hline
\end{tabular}
\label{tab_log1}
\end{table*}

\section{GRB\,070707}
\label{sec:grb}

\object{GRB\,070707} was detected in the 15--200 keV band with IBIS/ISGRI on
board the INTEGRAL satellite \citep{Me03} on 2007 Jul 7 at 16:08:21 UT, and
initially classified as a long event \citep{Be07}. Subsequent analysis of the
IBIS/ISGRI data determined a $2\farcm1$ accurate position centered at
$\mbox{RA(J2000)} = 17^{\rm h} 51^{\rm m} 00\fs14$, $\mbox{Dec(2000)} =
-68\degr 54\arcmin 51\farcs8$, and revealed that the burst consisted of a
single spike lasting about $1.1$\,s \citep{Go07}. A refined, complete analysis
of the INTEGRAL data of GRB\,070707 has been presented by \citet{McG08},
showing that its properties are all consistent with those of short GRBs: short
duration (0.8~s), very small spectral lags (20~ms), and hard spectrum (photon
index $\Gamma = 1.2$). Konus-Wind also detected this burst \citep{Gol07},
allowing the measurement of the broad-band fluence ($\sim 1.4 \times 10^{-6}$
erg cm$^{-2}$ in the 20\,keV--2\,MeV band) and peak energy ($\sim 400$\,keV).

The \textit{Swift} satellite began to observe \object{GRB\,070707} on 2007 Jul
08 at 00:57:50 UT, i.e. about 9\,hr after the IBIS/ISGRI trigger.
\textit{Swift}/UVOT did not detect any optical afterglow down to a $3\sigma$
limiting magnitude of $V = 19.7$ \citep{Sch07}. \textit{Swift}/XRT found a
faint, uncatalogued source inside the INTEGRAL error box, at $\mbox{RA(J2000)}
= 17^{\rm h} 50^{\rm m} 58\fs49$, $\mbox{Dec(J2000)} = -68\degr 55\arcmin
27\farcs1$ (positional uncertainty of $5\farcs4$), with a 0.3--10\,keV flux of
$2.4^{+2.0}_{-1.4} \times 10^{-13}$\,erg\,cm$^{-2}$\,s$^{-1}$ \citep{Bea07a}.
Further \textit{Swift}/XRT observations, carried out about 4.9\,days after the
burst, could not detect the source down to a ten times lower flux level. This
confirmed that the \textit{Swift}/XRT source was indeed the afterglow of
\object{GRB\,070707} \citep{Bea07b}.

ESO-VLT observations, carried out starting about half a day after the burst,
revealed the presence of a variable $R$-band source inside the
\textit{Swift}/XRT error circle, at $\mbox{RA(J2000)} = 17^{\rm h} 50^{\rm m}
58\fs55$, $\mbox{Dec(J2000)} = -68\degr 55\arcmin 27\farcs2$ ($0\farcs3$ error;
\citealt{Pi07,PDA07b}). No further observations of the afterglow of
\object{GRB\,070707} have been reported so far.

\section{Observations and data analysis}
\label{sec:obs}

We observed \object{GRB\,070707} with the ESO-VLT at eight different epochs
starting about $11$\,hr after the burst. Observations were carried out using
the FORS1, ISAAC and NACO cameras. All nights were clear, with seeing in the
0.5\arcsec--1.0\arcsec range. Image reduction was carried out following
standard procedures: subtraction of the bias frame and division by the flat
frame. Point spread function (PSF) and aperture photometry were obtained by
using the Daophot\,II \citep{Ste87} in the
ESO-MIDAS\footnote{\texttt{http://www.eso.org/projects/esomidas/}} package  for
all objects in the field. Photometric calibration was based on Landolt standard
stars, observed in different nights. In order to minimize systematic effects,
we performed differential photometry with respect to a selection of local
isolated and non-saturated standard stars. Astrometric solutions were computed
by using the USNO-B1.0
catalogue\footnote{\texttt{http://www.nofs.navy.mil/data/fchpix/}}. 

All our VLT/FORS spectra were acquired with the 300V grism, covering the
4000--9000\,\AA{} wavelength range (7\,\AA{} FWHM resolution). We always used a
1\arcsec{} slit. The extraction of the spectra was performed with the
IRAF\footnote{IRAF is distributed by the National Optical Astronomy
Observatories, which are operated by the Association of the Universities for
Research in Astronomy, Inc., under cooperative agreement with the National
Science Foundation.} software package. Wavelength and flux calibration of the
spectra were obtained by using helium-argon lamp and observing
spectrophotometric stars. A complete log of our observations, together with the
results of our analysis, is reported in Table\,\ref{tab_log1}.

\section{Results}
\label{sec:res}

\subsection{VLT photometry}

\begin{figure}
\includegraphics[width=\columnwidth]{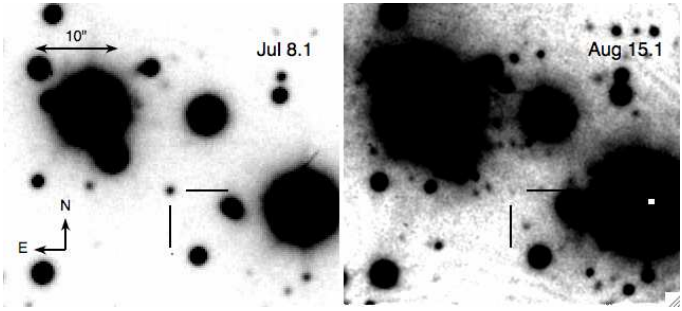}
\caption{$R$-band image of the optical afterglow ({\it left panel}) and of the
host galaxy ({\it right panel}) of \object{GRB\,070707}.}
\label{fig:picz}
\end{figure}

The optical light curve appears to have had an initial slow decay, which got
significantly steeper beginning 1--2 days after the GRB. At late times, a
constant flux was observed, indicating a dominant contribution from the host
galaxy (Fig.\,\ref{fig:picz}). We fitted the light curve with a model
comprising a broken power law behaviour, representing the afterglow, plus a
constant, representing the host. We adopted the usual Beuermann function
\citep{Beuermann99} to model a transition from a shallow to a steeper decay:
$F(t) = F_0 / [(t/t_{\rm b})^{\kappa\alpha_1} + (t/t_{\rm
b})^{\kappa\alpha_2}]^{1/\kappa}$, where $\alpha_1$ and $\alpha_2$ are the
early- and late-time decay slopes, $t_{\rm b}$ is the break time, and $\kappa$
is the smoothness parameter. In the case of GRB\,070707, given the relatively
low number of data points, we first froze $\kappa = 1$ in our fit. In
Table\,\ref{tab:fit} we report the best-fit parameters.

\begin{table}
\caption{Results of the optical-light curve fitting with a Beuermann function \citep{Beuermann99} freezing the smoothness parameter $\kappa = 1$. Errors are at $1\sigma$ with all parameters free to vary. }
\label{tab:fit}
\centering
\begin{tabular}{ccccc}
\hline
$\alpha_1$             & $\alpha_2$          & $t_{\rm b}$ (days)     & Host $R$ magnitude    & $\chi^2$/dof \\
\hline
$0.44^{+0.08}_{-0.21}$ & $5.3^{+0.9}_{-0.8}$ & $1.82^{+0.13}_{-0.25}$ & $27.3 \pm 0.13$ & 1.96/4       \\
\hline
\end{tabular}
\end{table}

\begin{table}
\caption{Results of the optical-light curve fitting with the ``pulse function''. Errors are at $1\sigma$ with all parameters free to vary. }
\label{tab:fit1}
\centering
\begin{tabular}{cccc}
\hline
$\tau$ (days)          & $t_0$ (days)            & Host $R$ magnitude    & $\chi^2$/dof \\
\hline
$0.52^{+0.07}_{-0.07}$ & $-0.06_{-0.29}^{+0.23}$ & $27.3 \pm 0.24$ & 2.28/5       \\
\hline
\end{tabular}
\end{table}

When $\kappa$ was allowed to vary, the break time was only weakly constrained
and the slope $\alpha_1$ could also take negative values (i.e. the light curve
could have also be initially rising and peak around 0.5 days). In any case the
late time decay remained steep, with a lower limit $\alpha_2 \ge 3$ valid for
any $\kappa$.

\begin{figure}
\includegraphics[width=\columnwidth]{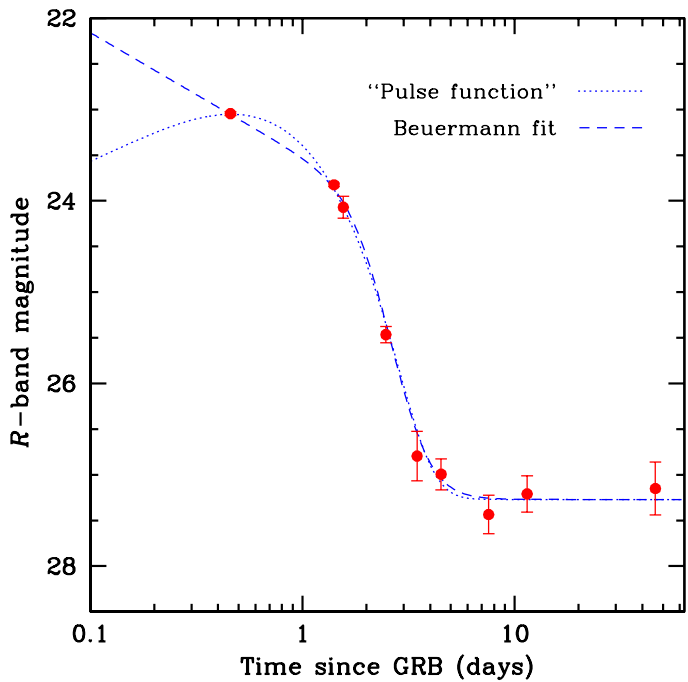}
\caption{$R$-band light curve of the GRB\,070707 afterglow. The dashed line
represents the fit obtained using a Beuermann function (with $\kappa = 1$),
while the dotted line shows the fit with the ``pulse function'' (see text).}
\label{lc}
\end{figure}

We note that different functional forms cannot be excluded for the light curve,
given the sparse coverage at early times. For example a model consisting of a
linear rise followed by an exponential decay $F(t) = F_1 (t-t_0)
e^{-(t-t_0)/\tau}$ (we refer to this as the ``pulse function'') provided a
satisfactory fit to the data (even for $t_0 = 0$, i.e. the origin of time
coincident with the high-energy event; Table \ref{tab:fit1}). The best fit
models with the Beuermann function (with $\kappa = 1$) and the ``pulse
function'' are shown in Fig.\,\ref{lc}.

\subsubsection{VLT spectroscopy}

After the identification of the optical counterpart  \citep{Pi07} we took an
optical spectrum of the afterglow (about 1.7 days after the burst; see
Table\,\ref{tab_log1}), consisting of five frames with a total exposure time of
200\,min. Due to the rapid decay of the afterglow, the object was clearly
visible only in the first two frames. As a consequence the total spectrum has a
mediocre signal-to-noise ratio. Neither emission nor absorption lines are visible
over a weak continuum in the 4200--9000 \AA{} range. The detection of the
continuum, on the other hand, allows us to put some constraints on the redshift
of this event. The most conservative limit on the redshift is given by the lack
of Lyman limit suppression down to 4200 \AA, which implies $z < 3.6$. The lack
of observed Lyman-$\alpha$ forest provides a tighter constraint $z \la 2.5$,
although at such redshift the density of the IGM is relatively low, making this
limit less robust.


\begin{table}
\caption{Properties of five galaxies in the field of GRB\,070707 (see Fig.\,3
and Sect.\,4 for details).}
\centering          
\begin{tabular}{ccccc}\hline
ID         & RA (J2000)   & Dec (J2000) & $R$ magnitude &  Redshift \\ \hline
G1         & 17:50:58.55  & -68:53:35.0 &  21.1       &  0.3547   \\
G2         & 17:50:58.40  & -68:53:01.5 &  20.5       &  0.3780   \\
G3         & 17:51:04.73  & -68:54:27.0 &  20.2         &  0.3627   \\
G4         & 17:51:00.67  & -68:55:01.8 &  17.6         &  0.213    \\
G5         & 17:50:58.97 & -68:55:12.8 &   20.5          &  0.667  \\
G6         & 17:50:57.18  & -68:55:29.4 &  20.4         &  0.2394   \\ \hline
\end{tabular}
\label{tab_z}
\end{table}

\begin{figure}
\includegraphics[width=\columnwidth]{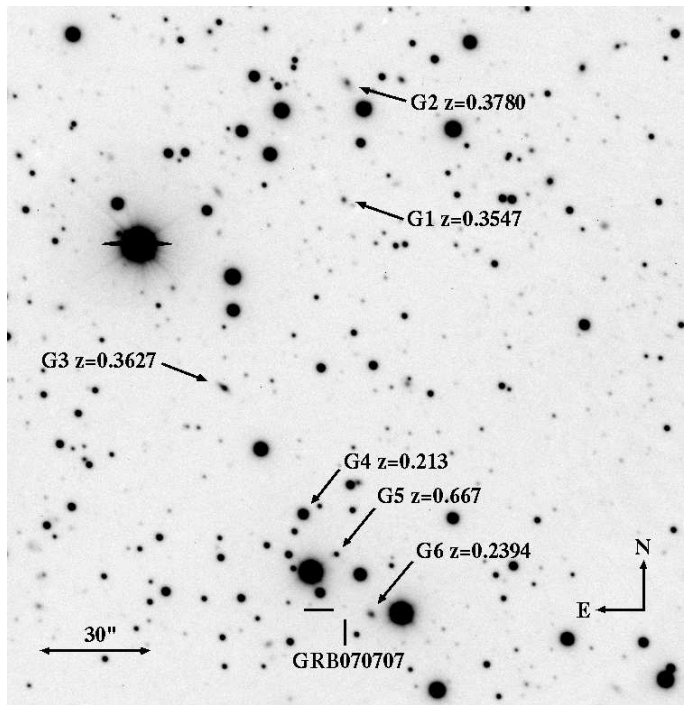}
\caption{$R$-band image of the field of \object{GRB\,070707}. The galaxies
whose redshifts have been reported in Table~\ref{tab_z} are marked, together to
the position of GRB\,070707. The field is about $3\arcmin \times 3\arcmin$
wide.}
\label{fig:z}
\end{figure}

\begin{figure}
\includegraphics[width=\columnwidth]{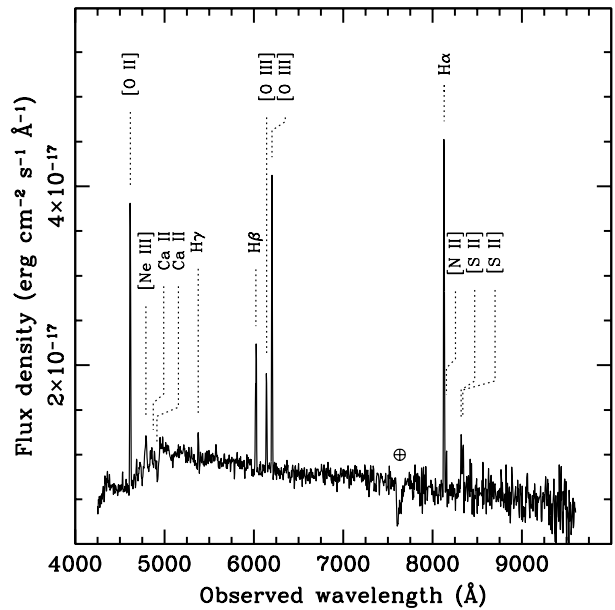}
\caption{VLT-FORS2 spectrum of galaxy G6 (see Fig.\,\ref{fig:z}) at $z =
0.2394$.}
\label{fig:G6}
\end{figure}

Many galaxies (a few tens) are clearly visible in our VLT images of
\object{GRB\,070707}. In order to check for the possible presence of a cluster
we obtained optical spectra of five galaxies in the field and determined their
redshifts (Table\,\ref{tab_z} and Fig.\,\ref{fig:z}). These spectra were taken
on 2007 Jul 9th and 13th (Table\,\ref{tab_log1}). The redshift gap between G1,
G2 and G3  (0.3547--0.3780) is very large and corresponds to a  velocity
difference of 7000\,km\,s$^{-1}$ with a dispersion of 3500\,km\,s$^{-1}$. This
value is considerably higher than those measured in clusters of galaxies;
therefore we consider very unlikely that these galaxies belong to a cluster.
The object labelled as G6 is the closest to the position of the GRB, with a
separation of 7\farcs8. A qualitative analysis of the optical spectrum of
galaxy G6 (Fig.\,\ref{fig:G6}) reveals that it is a starburst galaxy at $z =
0.2394$, as indicated by the presence of prominent nebular lines. If the GRB
were also occurring at this redshift, the projected offset would be about
29.3\,kpc, and its host galaxy would have an absolute visual magnitude of $M_V
\sim -13.2$, a value similar to dwarf galaxies of the Local Group, such as
Fornax, but still much brighter than a globular cluster.

\section{Discussion}
\label{sec:disc}

\begin{figure}
\includegraphics[width=\columnwidth]{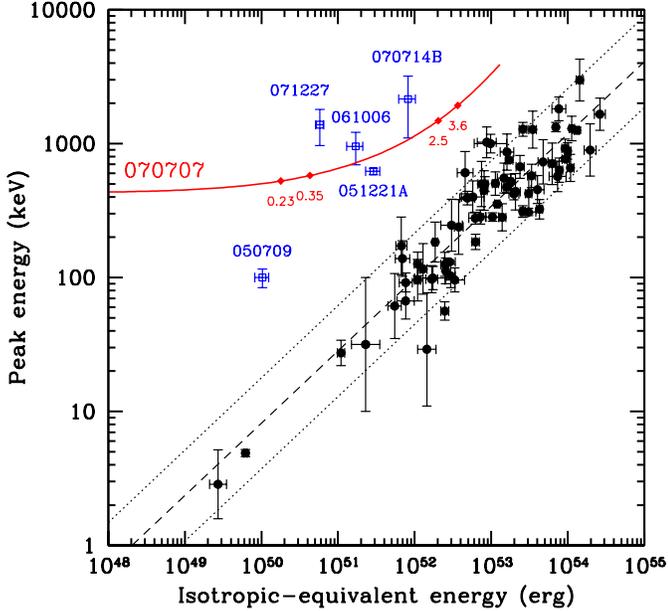}
\caption{Location of GRB\,070707 in the plane peak energy vs.
isotropic-equivalent energy. The thick solid line shows the position of
GRB\,070707 as a function of redshift, with the diamonds indicating specific
values discussed in this paper. Filled circles represent long-duration GRBs
(from \citealt{Amati08}), and the diagonal lines indicate the best-fit Amati
relation (dashed) and the 2$\sigma$ contours (dotted). Empty squares indicate
other short-duration events with known redshift and peak energy
\citep{Ama06,Gole06,Ohno07,Gole07b}.}
\label{Amati}
\end{figure}

As with a number of other short GRBs, the nearly unconstrained redshift of
GRB\,070707 remains an important limiting factor. In the peak energy vs.
isotropic-equivalent energy diagram (Fig. \ref{Amati}), \object{GRB\,070707} 
should lie on the thick solid curve marked with selected values of the redshift.
For the range of possible redshifts, the position of \object{GRB\,070707} would
be in broad agreement with that of other short GRBs with known redshift and
peak energy, which notoriously do not follow the so-called ``Amati-relation''
of long GRBs \citep{Ama06}.

Whichever its redshift, the host of \object{GRB\,070707} ($R \sim 27.3$) is the
faintest ever detected for a short GRB, with a magnitude comparable to that of
long GRB hosts at high redshift (e.g. \citealt{Wain07,Fruchter06}). In past
works \citep[e.g.][]{Blo07,Str07,Lev07}, there has been discussion about the
possibility to find short GRBs not spatially coincident with a host galaxy. In
some cases, nearby, bright objects were proposed to be the GRB host based just
on angular proximity. The case of GRB\,070707 shows that caution is needed.
Without deep VLT images, it might have been tempting to associate GRB\,070707
with the closeby galaxy G6, which is certainly not the GRB host. With such a
large magnitude contrast between the afterglow and the host, many more galaxies
of short GRBs may be just fainter than the fluxes probed by shallower
exposures. While this is consistent with the suggestion that a sizeable
fraction of short GRBs reside at redshift larger than $z \sim 1$
\citep{Ber07b}, this may also indicate that some short GRBs go off inside
low-luminosity galaxies at low redshift.


The sparse data on the host galaxy of GRB\,070707 do not allow for a detailed
analysis. At the maximum allowed redshift $z = 3.6$, the host would have an
absolute magnitude $M_{\rm AB} = -18.6$ (at rest-frame wavelength $\lambda
\approx 1500$~\AA). This is more that 2 magnitudes fainter than the Schechter luminosity at this redshift ($M^*_{\rm AB} = -20.7$; \citealt{Gabasch04}).
We can thus firmly set $L < (1/6)L_*$ at any redshift, implying that the host was intrinsically faint.
Our late-time NIR upper limit ($K > 22.7$) implies $R - K < 4.3$ and hence can
rule out a bright, red host, such as an extremely red object (ERO) or a
moderate-redshift elliptical. GRB\,070707 hence confirms that short GRBs can
explode inside faint and possibly extremely faint systems. Short GRB hosts
indeed exhibit a wide range of luminosities.

Our spectrum did not show any clear absorption feature, which is quite possibly
due to the low signal-to-noise and/or limited covered wavelength range. It is
however interesting to note that a featureless spectrum has been reported also
for the afterglow of the short GRB\,061201 \citep{Str07}. If confirmed by
further, better-quality observations, this fact may provide hints for a lower
density of short GRB environments compared to those of long-duration events.

Further information on the GRB progenitor comes from the optical light curve of
the GRB\,070707 afterglow, which decayed very steeply starting $\sim 1.5$~days
from the burst. The power law decay index $\alpha$ is constrained by our fits
to be steeper than 3, or may even be exponential. Assuming that the optical
afterglow emission came from the forward shock, as commonly supposed in the
fireball model for GRB afterglows, the steepening of the optical light curve
could be interpreted as a jet break. However the index measured for GRB\,070707
is too steep to be explained in terms of jetted emission. In fact this would
require $\alpha_2 = p$ \citep{Rhoads99}, where $p$ is the electron energy
distribution index. Such a steep decay could be marginally consistent with the
post-break phase only by adopting a very soft electron energy distribution ($p
> 3$). Such large $p$ values have never been found in GRB afterglows, both from
theoretical and empirical investigations (e.g. \citealt{Pana05,She06,Tag06, Kann06}).
In particular, values of $p < 3$ are inferred based on afterglow spectra, in
which case the analysis is more robust than when relying on the temporal
behaviour. Moreover, the magnitude of the steepening from the pre- to the
post-break decay would be too pronounced for a jet-break interpretation in
GRB\,070707.


The steepness of the  observed decay might be difficult to reconcile with
forward shock emission even just for causality reasons. The fastest possible
decay is the so-called high-latitude emission, which occurs when the fireball
emission stops abruptly and the observers see photons coming from the wings of
the emitting surface. In this case, $\alpha = 2 + \beta$, where $\beta$ is the
afterglow spectral index \citep{Kumar00}. The observed upper limit in the $J$
band (Table\,\ref{tab_log1}) allows us to derive  $R-J < 2.1$ on Jul 10.2 UT,
which corresponds to $\beta < 1.95$. This would still be consistent with the high-latitude interpration. 
A stronger limit on $\beta$ can however be inferred by using the X-ray data. Assuming a
synchrotron spectrum, the optical spectral index can never be softer than the
optical-to-X-ray spectral index $\beta_{\rm OX}$. For GRB\,070707, the observed
X-ray flux \citep{Bea07b} implies $\beta_{\rm OX} = 0.75^{+0.13}_{-0.09}$. High-latitude
emission from a source with such a spectrum cannot decay faster than
$t^{-2.75}$, hence effectively ruling out such possibility for GRB\,070707.

%

%

A viable alternative to explain the steep decay of GRB\,070707 is that of a
long-lived central engine. In this case, as discussed by several authors
\citep[e.g.][]{Zhang06}, the decay index should be computed after setting the
zero time $t_0$ to the end of the extended activity phase. The intrisic decay slope would therefore be shallower than the observed value, eliminating the causality problem.
The very steep decay of \object{GRB\,070707} thus would provide further evidence that the inner
engine powering short GRBs is working for a much longer time than the observed
gamma-ray emission. The optical light curve of \object{GRB\,070707} might be
produced, for example, by a large flare, as also argued for
\object{GRB\,050724} \citep{Barthelmy05,Malesani07}. 

Finally, we mention that the optical light curve of GRB\,070707 could also be explained within the context of cannonball model (\citealt{Dado08a}; \citealt{Dado08b} and references therein). The emitting region in this case has angular dimension well below $1/\Gamma$, and therefore the causality constraints are much more relaxed.

\begin{acknowledgements}
We thank an anonymous referee for comments. We thank D. A. Kann and A. Dar for useful discussion. 
Part of this work was supported by MIUR COFIN-03-02-23 and INAF/PRIN 270/2003
and ASI contracts ASI/I/R/039/04 and ASI/I/R/023/05/0. S.D.V. is supported by
SFI. The Dark Cosmology Centre is funded by the Danish National Research
Foundation. APB is supported by STFC.
\end{acknowledgements}

\end{document}